\documentclass[11pt,twoside]{article}
\usepackage{asp2006}
\usepackage{epsf}
\usepackage{graphicx}
\usepackage{lscape}
\def\edcomment#1{\iffalse\marginpar{\raggedright\sl#1\/}\else\relax\fi}
\reversemarginpar

\begin{document}
\title{Signatures of resonant terrestrial planets in long-period systems}
\author{Gareth F. Kennedy}
\affil{Institut de Ci\`{e}ncies del Cosmos,Universitat de Barcelona, 08028 Barcelona, Spain}
\author{Rosemary A. Mardling}
\affil{School of Mathematical Sciences,Monash University, Clayton 3800, Australia}

\begin{abstract}
The majority of extrasolar planets discovered to date have significantly eccentric orbits, some if not all of which may have been produced through planetary migration. During this process, any planets interior to such an orbit would therefore have been susceptible to resonance capture, and hence may exhibit measurable orbital period variations. Here we summarize the results of our investigation into the possibility of detecting low-mass planets which have been captured into the strong 2:1 resonance. Using analytical expressions together with simulated data we showed that it is possible to identify the existence of a low-mass companion in the internal 2:1 resonance by estimating the time-dependant orbital period for piecewise sections of radial velocity data. This works as long as the amplitude of modulation of the orbital period is greater than its uncertainty, which in practice means that the system should not be too close to exact resonance. 
Here we provide simple expressions for the libration period and the change in the observed orbital period, these being valid for arbitrary eccentricities and planet masses. They in turn allow one to constrain the mass and eccentricity of a companion planet if the orbital period is sufficiently modulated.
\end{abstract}

\vspace{-0.5cm}
\section{Resonance theory}

The following summarizes the results of Kennedy \& Mardling (in preparation).
A resonant system is characterized by the libration of one or more resonance angles. For a coplanar system the quadrupole 2:1 resonance angle is given by
\begin{equation}
\phi=\lambda_i-2\lambda_o+\varpi_i
\end{equation}
where $\lambda_i$ and $\lambda_o$ are the inner and outer mean longitudes, and $\varpi_i$ and $\varpi_o$ are the corresponding longitudes of periastron.
The maximum possible variation in $\sigma=\nu_i/\nu_o$, the ratio of inner to outer orbital frequencies, is given by
 (Mardling 2008)
\begin{equation}
\Delta\sigma=3\left[f(e_i)\,g(e_o) \left((m_o/m_*)+2^{2/3}(m_i/m_*) \right)  \right]^{1/2},               
\label{eqnOne}
\end{equation}
where
$e_i$ and $e_o$ are the inner and outer orbital eccentricities, $m_i$ and $m_o$ are the corresponding planet masses,\begin{equation}
f(e_i)= 3e_i-\frac{13}{8}e_i^3-\frac{5}{192}e_i^5+{\cal O}(e_i^7)
\hspace{0.3cm}{\rm and}\hspace{0.3cm}
g(e_o)=1-\frac{5}{2}e_o^2+\frac{13}{16}e_o^4+{\cal O}(e_o^6).
\end{equation}
For stable systems inside the 2:1 resonance the libration period is approximately
\begin{equation}
P_{lib}=2P_o/\Delta\sigma=\alpha [f(e_i)]^{-1/2}P_o,
\label{plib}
\end{equation}
where $\alpha$ depends on observed parameters when $m_i \ll m_o$. 
The amplitude of variation of the outer (observed) orbital period depends on the distance from exact resonance,
$\delta\sigma$, and is given by
\begin{equation}
\delta P_o/P_o=-2\left[1+(m_o/m_i)\sigma^{-2/3}\right]^{-1} \delta\sigma/\sigma\simeq -2^{2/3}\,(m_i/m_o)\,\delta\sigma,
\label{delpo}
\end{equation}
where $\delta\sigma=\sigma-2$.
If the period and amplitude of variation of the orbital period of a known system can be measured, (\ref{plib})
and (\ref{delpo}) can be used to estimate $e_i$, $m_i$ and $\delta\sigma$.

\section{Data Analysis}

\begin{figure}[!ht]
\begin{center}
\includegraphics[height=3.57cm]{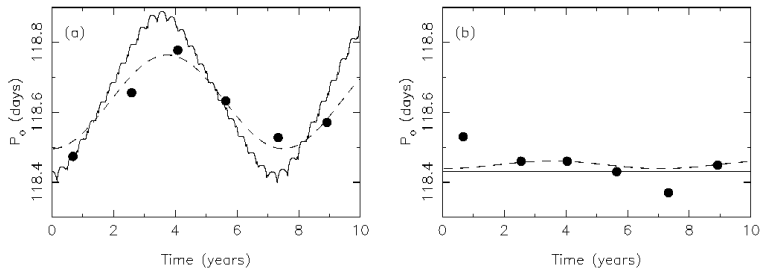}
\caption{Piecewise estimates (filled circles) of the orbital period for a simulated HD 216770-type system (a) with and (b) without a $10 M_\oplus$ companion in the interior 2:1 resonance. The solid curves are  from three-body integrations while the dashed curves are sine curve fits to the points, constrained by knowledge of possible values for the libration period. Simulated radial velocity data includes instrument and stellar jitter noise.}
\label{PTPlot}
\end{center}
\end{figure}

To illustrate our procedure we used the orbital parameters of HD 216770b for which
$m_o \sin i = 0.65\, M_J$, $M_* = 0.9 M_\odot$, $e_o = 0.37$, $P_o = 118.45$ days and $P_{lib}\simeq 25 P_o$.
Using synthetic data for a system with and without a 10 Earth-mass planet in the interior 2:1 resonance,
we were able to infer the presence of the super-Earth as follows.
First we determined the range of values of $e_i$ for which stable orbits exist using direct three-body integrations.
Sampling at the rate of around 16 synthetic radial velocity data points per orbit and five orbits per segment, 
we found single-orbit solutions to each segment
using a Levenberg-Marquardt algorithm. The resulting successive orbital periods are then fit by a sinusoid
and the period and amplitude compared with (\ref{plib}) and (\ref{delpo}). Using this procedure we were able
to recover the true libration period and amplitude to within 30\%. This is illustrated in Figure~\ref{PTPlot}.

\end{document}